\documentclass[sigconf, screen, authorversion]{acmart}
\usepackage{booktabs}
\usepackage{tabularx}
\usepackage{hhline}
\usepackage{graphicx}
\usepackage{cleveref}
\usepackage{metalogo}
\usepackage{verbatim}
\usepackage{listings}
\usepackage{csquotes}
\usepackage{tablefootnote}
\usepackage{float}
\usepackage{dblfloatfix}

\makeatletter
\renewcommand\@formatdoi[1]{\ignorespaces}
\def\@copyrightspace{\relax}
\makeatother

\crefformat{footnote}{#2\footnotemark[#1]#3}

\setcopyright{none}
\ccsdesc[500]{Applied computing~Markup languages}
\ccsdesc[500]{Applied computing~Document metadata}
\ccsdesc[500]{Applied computing~Annotation}

\acmConference[JCDL 2023]{ACM/IEEE Joint Conference On Digital Libraries}{June 26 - 30, 2023}{Santa Fe, New Mexico, USA}
\acmYear{2023}
\copyrightyear{2023}

\begin{document}
	%
	%\metatitle{\title{SciKGTeX - A \LaTeX{} Package to Semantically Annotate Contributions in Scientific Publications}}
	\title{SciKGTeX - A \LaTeX{} Package to Semantically Annotate Contributions in Scientific Publications}
	\renewcommand{\shorttitle}{SciKGTeX}
	\author{Christof Bless}
	\orcid{0000-0001-9778-8495}
	\affiliation{%
		\institution{Lucerne University of Applied Sciences}
		\streetaddress{Werftestrasse 4}
		\city{Lucerne}
		\country{Switzerland}
	}
	\email{christofbless@gmail.com}
	
	\author{Ildar Baimuratov}
	\orcid{0000-0002-6573-131X}
	\affiliation{%
		\institution{L3S Research Center\\ Leibniz University Hannover}
		%\streetaddress{P.O. Box 1212}
		\city{Hannover}
		\country{Germany}
		%\state{Ohio} 
		%\postcode{43017-6221}
	}
	\email{baimuratov.i@gmail.com}
	
	\author{Oliver Karras}
	\orcid{0000-0001-5336-6899}
	\affiliation{%
		\institution{TIB - Leibniz Information Centre for Science and Technology}
		%\streetaddress{P.O. Box 1212}
		\city{Hannover}
		\country{Germany}
		%\state{Ohio} 
		%\postcode{43017-6221}
	}
	\email{oliver.karras@tib.eu}
	
	%\metaauthor*{\uri{https://orcid.org/0000-0001-9778-8495}{Christof Bless}}
	%\metaauthor*{\uri{https://orcid.org/0000-0002-6573-131X}{Ildar Baimuratov}}
	%\metaauthor*{\uri{https://orcid.org/0000-0001-5336-6899}{Oliver Karras}}
	%\researchfield*{\uri{https://orkg.org/resource/R278}{Information Science}}
	
	\begin{abstract}
		%The continuously increasing output of published research makes the work of researchers harder as it becomes impossible to keep track of and compare the most recent advances in a field.
		Scientific knowledge graphs have been proposed as a solution to structure the content of research publications in a machine-actionable way and enable more efficient, computer-assisted workflows for many research activities. 
		%research problem
		%\researchproblem*{crowd-sourcing for scientific knowledge graphs}
	    Crowd-sourcing approaches are used frequently to build and maintain such scientific knowledge graphs.
		%Researchers are motivated to contribute to these crowd-sourcing efforts as they want their work to be included in the knowledge graphs and benefit from applications built on top of them.
		To contribute to scientific knowledge graphs, researchers need simple and easy-to-use solutions to generate new knowledge graph elements and establish the practice of semantic representations in scientific communication. %, there need to be simple and easy-to-use solutions to integrate the annotation process into the researcher's workflow.
		%objective
		In this paper, we present a 
		%\objective{workflow for authors of scientific documents to specify their contributions}
		workflow for authors of scientific documents to specify their contributions with a \LaTeX{} package, called SciKGTeX, and upload them to a scientific knowledge graph.
		%\objective*{automatically upload contributions to a knowledge graph}
		automatically upload contributions to a knowledge graph
		%SciKGTeX, \objective{a \LaTeX{} package to semantically annotate scientific contributions} at the time of document creation.
		%method
		%\method*{latex}
		%\method*{luatex}
		The SciKGTeX package allows authors of scientific publications to mark the main contributions %such as the research problem, objective, method, results and conclusion 
		of their work directly in \LaTeX{} source files. The package embeds marked contributions as metadata into the generated PDF document, from where they can be extracted automatically and imported into a scientific knowledge graph, such as the ORKG.
		This workflow is simpler and faster than current approaches, which make use of external web interfaces for data entry.
		%Thus, \result{adding or updating contributions in the ORKG using SciKGTeX requires no manual input and can be achieved faster}. %Ildar
		%In addition to the package, we document a \method{user evaluation with 26 participants} which I conducted to assess the usability and feasibility of the solution.
		%result
		Our user evaluation shows that SciKGTeX is easy to use, with a 
		%\result{score of 79 out of 100 on the System Usability Scale}
		score of 79 out of 100 on the System Usability Scale, as participants of the study needed only 
		%\result{7 minutes on average to annotate the main contributions}
		7 minutes on average to annotate the main contributions on a sample abstract of a published paper.
		%Furthermore, the study showed that the functionalities of the package can be picked up very quickly by the study participants which only needed \result{7 minutes on average to annotate the main contributions} on a sample abstract of a published paper.
		%conclusion
		Further testing shows that the embedded contributions can be successfully uploaded to ORKG within ten seconds.
		%\conclusion{SciKGTeX simplifies the process of manual semantic annotation of research contributions in scientific articles. Our workflow demonstrates how a scientific knowledge graph can automatically ingest research contributions from document metadata.}
		SciKGTeX simplifies the process of manual semantic annotation of research contributions in scientific articles. Our workflow demonstrates how a scientific knowledge graph can automatically ingest research contributions from document metadata.
		
	\end{abstract}
	\keywords{Semantic Annotation, \LaTeX{}, FAIR data, Scientific Knowledge Graphs.}
	
	\maketitle              
	
	\section{Introduction}
\label{sec:introduction}
%% Motivation
%% Context
Scientific discoveries have long become a community effort, with sometimes hundreds of researchers from different institutions collaborating on solutions to increasingly complex research problems.
While problems and approaches in research have evolved greatly over the years, scientific communication has still a lot of potential to improve.
Nowadays, the standard process in scientific communication is to publish scientific articles which are archived and distributed as PDF files~\cite{johnson2018}.
%% Problem
This is a basic approach to digitizing research content and does not leverage modern technologies which could pave the way to computer-assisted knowledge exchange.
With the immense number of published articles, it gets increasingly harder to keep an overview of the state-of-the-art in certain fields while at the same time, reproducibility of research~\cite{aarts2015, baker2016} and quality of peer reviews have been stagnating~\cite{smart_reviews}.

%Ildar+
%% Scientific Knowledge Graphs as solution
As a possible solution to this problem, scientific knowledge graphs, such as the Open Research Knowledge Graph\footnote{\url{https://orkg.org/}} (ORKG)~\cite{auer2020, Stocker.2023}, have been developed~\cite{Stocker.2022}. 
%The ORKG is aimed to facilitate research by representing scientific contributions in a structured and semantic way, readable by machines and humans.
Scientific knowledge graphs represent research content in a graph network of relations and concepts, which allows more sophisticated methods of information extraction.
%Christof+
%Scientific knowledge graphs allow to find publications not just based on the choice of keywords which might or might not occur in the text.
%In a knowledge graph, search programs can access networks of terms and concepts which can be found in the articles' text.
Unlike raw text, graph networks contain a semantic representation of the content which is more structured and consistent.
Knowledge graphs lay a solid foundation for a plethora of applications which can exploit such semantically enriched graph structures. 
%% What are some examples
Among the possible applications are enhanced document retrieval techniques \cite{fensel2017}, automatic literature reviews \cite{Oelen2019}, reasoning engines, autonomous research systems \cite{Nikolaev2016}, mathematical proof assistants \cite{Lange2011} and paper recommendation systems \cite{Cuong2020,Medvet2014}.

\begin{figure}[htbp]
    \centering
    \includegraphics[width=\linewidth]{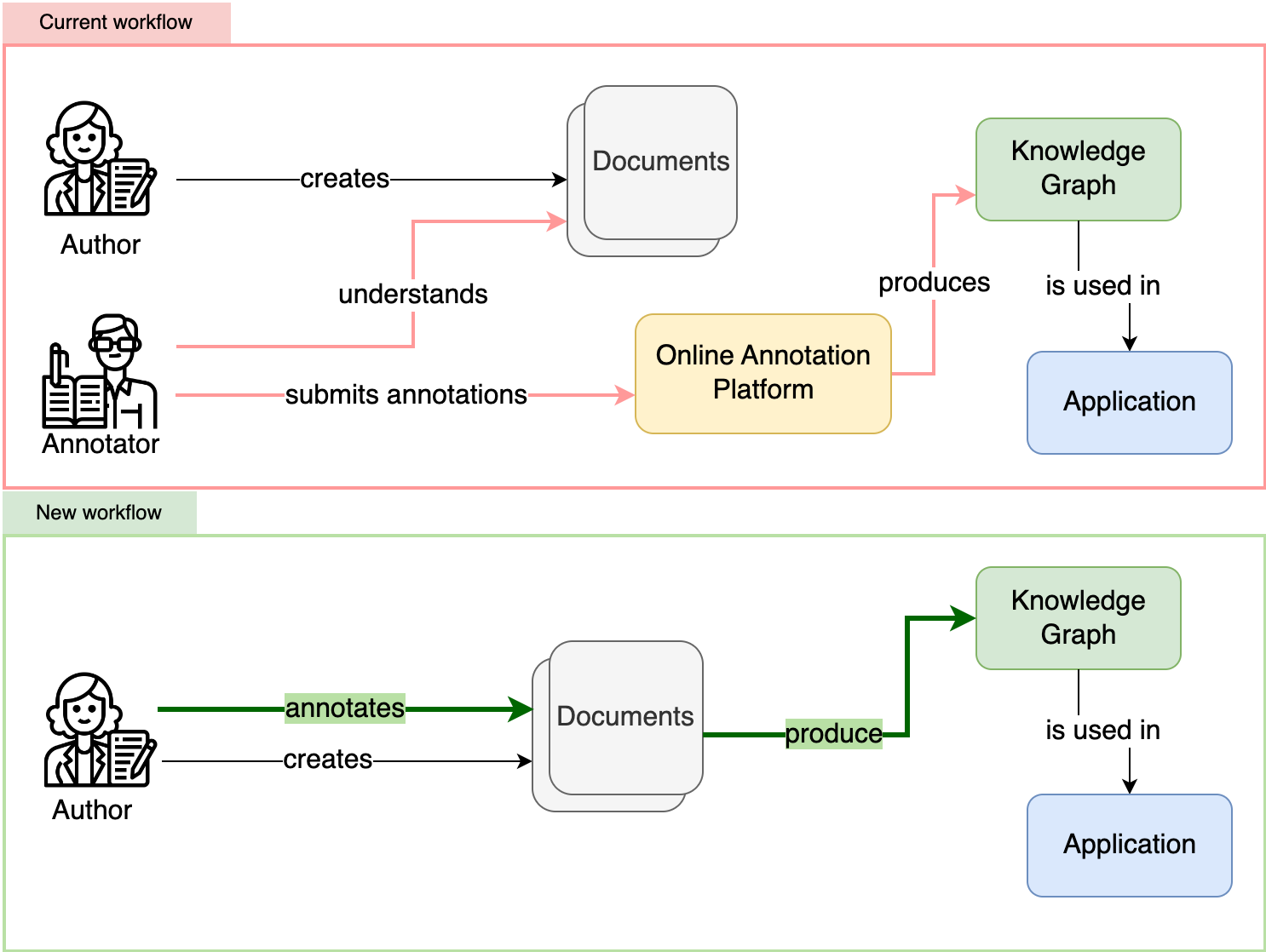}
    \caption{Our proposed solution simplifies the crowd-sourcing approach of scientific knowledge graphs.}
    \label{fig:idea}
\end{figure}

%% How are Sci KGs created?
The creation of complex, high quality knowledge graphs requires domain and ontology experts to define concepts and relations of the graph.
These concepts and relations must be identified in research texts and then extracted into the knowledge graph.
A common strategy to achieve annotation on a large scale is through crowd-sourcing \cite{takis2015}.
In the case of the ORKG, crowd-sourcing is realized through an annotation tool provided on a web platform~\cite{karras2021}.
%Christof-
%% Problem with Scientific Knowledge Graphs crowd-sourcing
The problem with this approach is that the annotation of concepts and relations is a laborious task which discourages potential users. Scientific contributions are already often fragmented across a multitude of platforms such as dataset repositories, preprint websites, postprint discussion threads and video platforms. Adding yet another platform to this mix distracts the researchers from their main objectives and complicates their workflow.
To incentivize researchers contributing to scientific knowledge graphs, the annotation of metadata needs to be integrated seamlessly into the scientific process. Figure \ref{fig:idea} illustrates the complexity problem of the current workflow (red).
%The process should be as simple and pain free as possible.
Furthermore, the metadata which the researchers generate should not end up in a data silo and possibly vanish with time. %establishing the practice of semantic representations in scientific communication,

%% Goal
%TODO more abstract description of the goal
%+Christof 

\begin{figure}[!t]
    \centering
    \includegraphics[width=\linewidth]{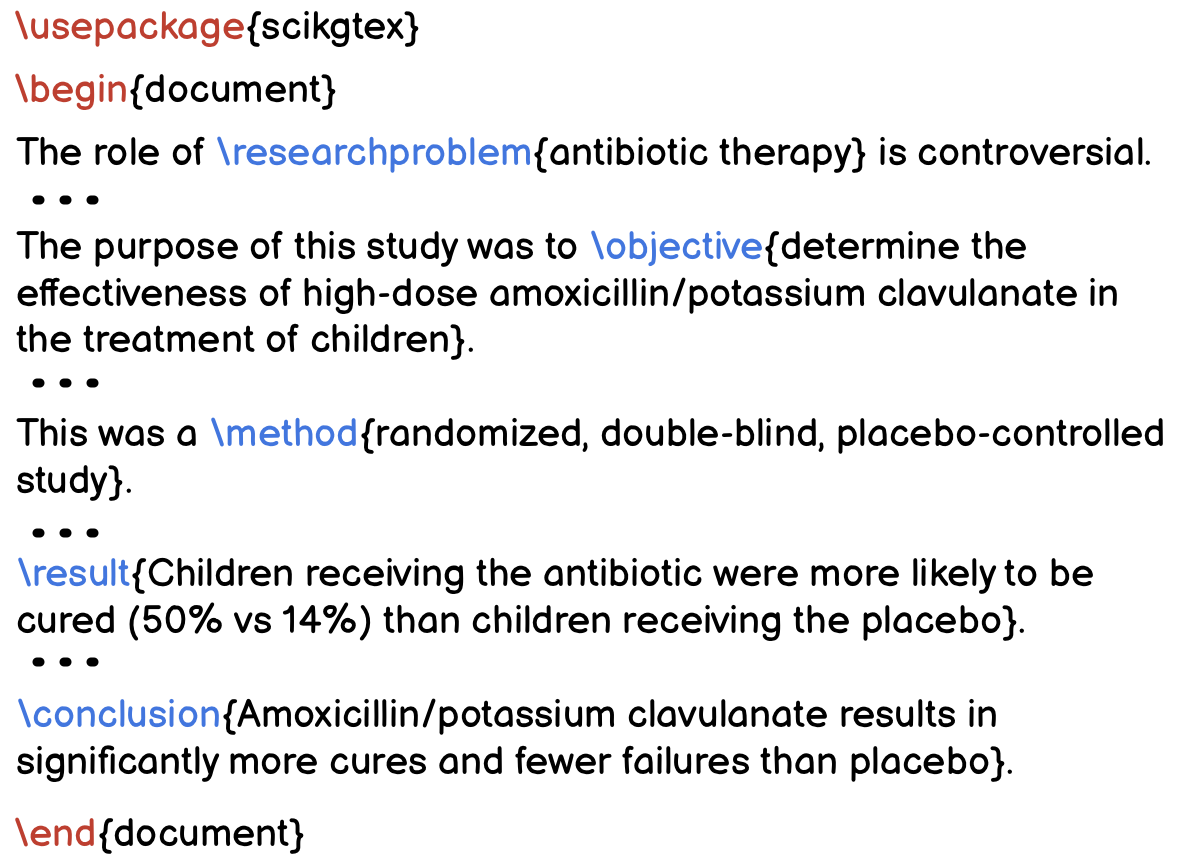}
    \caption{A possible application of the LaTeX package on a simplified version of a scientific article.}
    \label{fig:sketch}
\end{figure}

%% Research questions
We propose a new solution which utilizes commonly used systems and technologies in a researcher's toolbox. The gain in simplicity which can be achieved by our approach can be seen in Figure \ref{fig:idea} in the workflow highlighted in green. A part of the complexity of the current process is simplified by treating the annotation of metadata and the creation of the document as a common step. Our idea is to find a solution which does not rely on an online annotation platform or even a third-party annotator to submit the contributions. The system will integrate with existing metadata aggregation ecosystems as a metadata specification tool which enables authors to declare metadata pre-publication as opposed to post-publication. We ask the following research questions to develop our solution:
\begin{description}
    \item[RQ 1:] \hypertarget{RQ1}{How can the process of manual semantic annotation of research contributions in scientific articles be simplified?}
    %\paragraph{RQ 2} \textit{How do researchers use SciKGTeX to semantically annotate research contributions in scientific articles at the time of document creation?}
    \item[RQ 2:] \hypertarget{RQ2}{How can a scientific knowledge graph automatically ingest research contributions from document metadata?}
\end{description}
To answer RQ 1, we develop the SciKGTeX \LaTeX{} package as sketched in Figure~\ref{fig:sketch} and conduct a user evaluation which highlights our solution's simplicity and usability.
RQ 2 is about the import of PDF metadata to a scientific knowledge graph. We investigate the practicability of ingesting XMP metadata from PDFs into the ORKG.

%% Approach of the paper and our contribution
In this paper, we present a complete workflow of solutions for annotating, embedding, extracting, and importing structured research contributions from scientific documents typed in \LaTeX{}.
The workflow includes three components: i) the SciKGTeX \LaTeX{} package with markup capabilities for contributions, ii) the \LuaTeX{} PDF compiler, and iii) the PDF2ORKG import module, where
SciKGTeX and PDF2ORKG are the original contributions of this work.

The outline of the paper is the following: in Section 2 we overview the related work on semantic annotation in \LaTeX{}. In Section 3, we describe our approach through user stories and functionalities. Section 4 provides details on the implementation of the SciKGTeX package, and PDF2ORKG import module. Section 5 describes the user evaluation which we conducted on the system and reports the results of the user evaluation.
Finally, in Sections 6 and 7 we discuss the research questions and further implications of our work and draw the final conclusion.
	\section{Related Work}
\label{sec:related}
%SALT
Some of the first documented attempts to solve the problem of semantic metadata specification in \LaTeX{} were made in 2007 by Groza et al. \cite{Groza2007} who published a framework to semantically annotate structural and content-related text elements in \LaTeX{}.
The framework is called SALT (Semantically Annotated \LaTeX{}) and comprises a \LaTeX{} package with annotation commands and an annotation schema consisting of three ontologies.
Similar to the approach chosen in this work, the annotations are stored in the PDF metadata field, albeit the use case is slightly different.
They concentrate primarily on the generation of HTML content from the annotated PDF to support the automatic creation of online proceedings, but do not explore other use cases.
Moreover, SALT is not maintained anymore and cannot be used at the time of writing.

Moreau et al. \cite{moreau2015} released a \LaTeX{} package which can be used to add provenance information to a document. As ``provenance'' they define a record that describes how entities, activities and agents have influenced a piece of data.
Their package generates RDF statements for different types of provenance and saves them in a TURTLE file.
It also adds a link to the TURTLE file into the XMP metadata field of the PDF file but fails to embed the data itself into the PDF document.
Another semantic annotation markup was developed by Michael Kohlhase \cite{kohlhase2008} to turn \LaTeX{} into a document format for mathematical knowledge management (MKM).

%Ildar+
Most recently, Martin \& Henrich \cite{RDFtex} worked on a similar objective of linking \LaTeX{} publications with scientific knowledge graphs. They implemented the RDFTeX framework, which enables importing existing contributions from scientific knowledge graphs in \LaTeX{} and exporting new contributions in RDF format. However, the authors do not consider embedding the contributions into the PDF file and rather store them in an additional RDF document, which is detrimental to the persistence of the metadata.
On top of that, they do not show the exchange with an existing scientific knowledge graph and only describe it in theory.
We extend their work by providing a refined \LaTeX{} package which strives to be more intuitive for regular scientific writers, as it does not require the additional step of preprocessing the \LaTeX{} files with a Python script. Furthermore, we contribute a first user evaluation to back our usability claims 
 and set a benchmark for other similar tools.
%Ildar-

Another related solution for crowd-sourcing metadata annotation is the PDF annotation tool for the ORKG which is described by Oelen et al.~\cite{oelen2021}.
This tool can be used through a graphical user interface on the ORKG.
While this approach seems viable for published papers, we argue that adding the annotations directly in \LaTeX{} instead of first converting to PDF is a better approach since it is simpler and less error-prone.
For example, it is easier to annotate text which spans across pages, is formatted in tables or can not be extracted from PDF such as mathematical expressions.
Moreover, with our proposed workflow, it is not necessary for the authors to navigate to an annotation website and leave the usual \LaTeX{} working environment. This resonates with the results of our evaluation, which shows that SciKGTeX scores higher on a standardized usability test compared to the PDF annotation tool by Oelen
et al.~\cite{oelen2021} (see Section~\ref{sec:results}).

In general, it should be noted that most metadata initiatives concentrate on bibliographic metadata and do not provide ways to encode machine-actionable  representations of the actual scientific content of documents.
Our solution serves as a practical implementation of content-related metadata specification and storage, which is easier to use and adopt than previous approaches.

\section{Approach}
\begin{figure*}[!t]
    \centering
    \includegraphics[width=\linewidth]{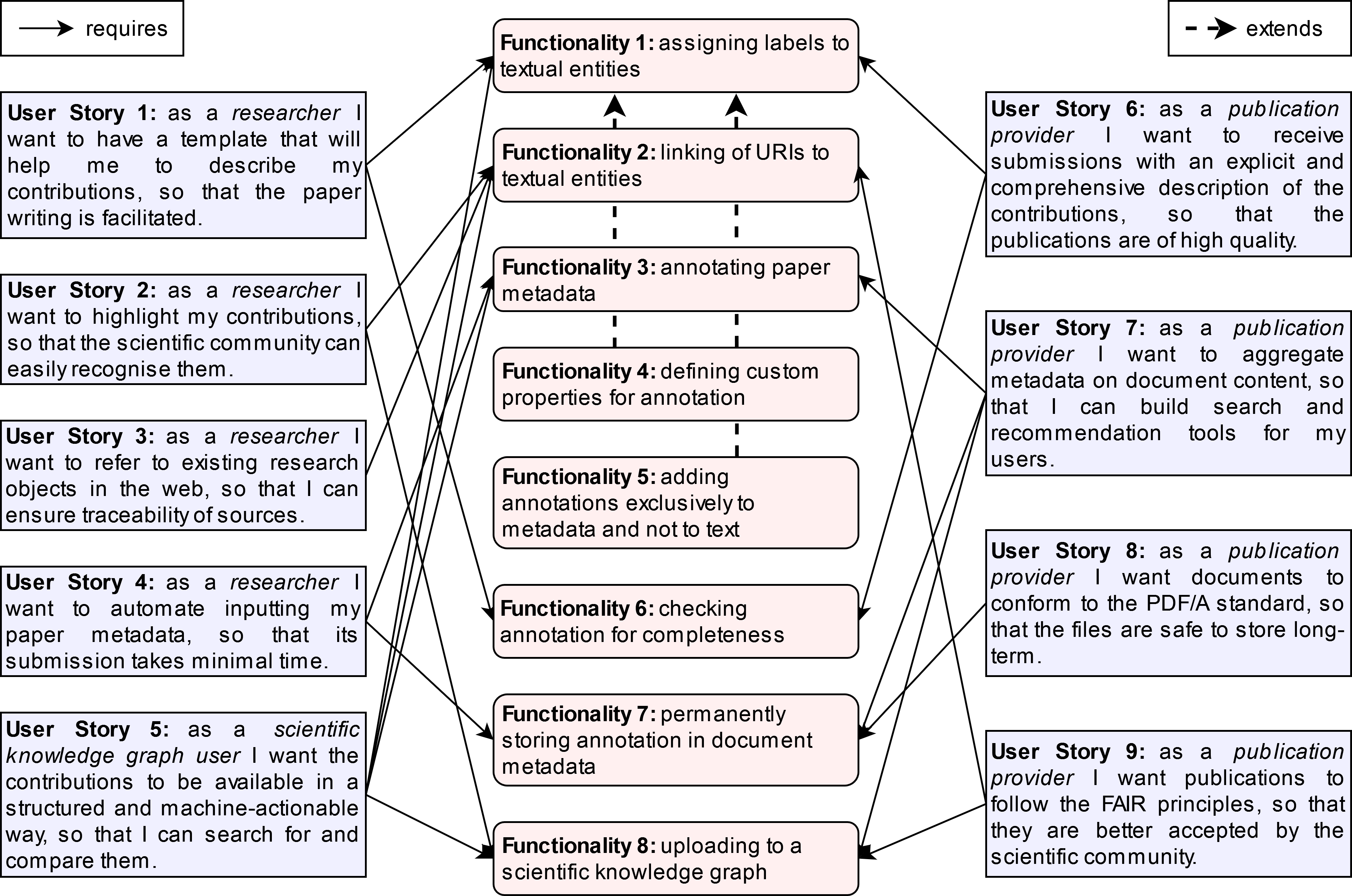}
    \caption{User stories and functionalities}
    \label{fig:stories}
\end{figure*}

For the development of the scientific knowledge graph annotation workflow described in Section \ref{sec:introduction}, we followed an agile development approach \cite{beck2001agile}.
The first step in our approach consisted of assessing the requirements that users have for the tool from a variety of possible use cases, which were determined from conversations with researchers.
The requirements are expressed as user stories \cite{cohn2004} following the \textit{Connextra} template \cite{lucassen2016}.
In the \textit{Connextra} template, a user story has 3 slots -- a role, a requirement and an optional reason.
The slots are connected into a sentence:

\vspace{0.1cm}
As a \textit{$\langle$role$\rangle$} I want to \textit{$\langle$requirement$\rangle$}, so that \textit{$\langle$reason$\rangle$}.
\vspace{0.1cm}
%\begin{displayquote}
%As a \textit{$\langle$role$\rangle$} I want to %\textit{$\langle$requirement$\rangle$}, so that %\textit{$\langle$reason$\rangle$}.
%\end{displayquote}

The different user stories are organized into three roles which were identified -- the researcher, the publication provider, and the scientific knowledge graph user.
The notion of researcher specifically stands for an author of a scientific publication here.
A publication provider denotes an entity or organization which collects and curates scientific publications and distributes them to the greater public.
This includes publishers of academic journals, conference proceedings and books as well as library services and archive platforms such as \textit{arXiv}\footnote{\url{https://arxiv.org/}}.
A scientific knowledge graph user operates with the structured and machine-actionable representation of scientific contributions provided by platforms such as the ORKG.
For example, it can be a literature review author who intends to use automated comparison platforms to supplement the creation of their review.

Afterward, we came up with specific functionalities which address the user stories. These functionalities were then implemented in the first prototype.
The lists of stories, functionalities, and their relationships are represented in Figure~\ref{fig:stories}. 

We also adopted the principle of iterative development cycles from the agile approach.
This means that after implementing the functionalities, we evaluated the resulting system either through a review process or the user evaluations described in Section \ref{sec:evaluation}.
Then, the cycle restarts with the definition of new user stories gained from the review process. 
	\section{Implementation}

%Ildar+
The proposed pipeline consists of two separate solutions: i) the SciKGTeX package for \LaTeX{}, and ii) the PDF2ORKG import module to showcase ingestion of the metadata into a centralized scientific knowledge graph.

\subsection{SciKGTeX Package}\label{sec:package}

To implement the specified functionalities (Fig. \ref{fig:stories}: Functionalities 1--7), we developed the SciKGTeX package for \LaTeX{}. \LaTeX{} is a popular tool for the creation of scientific publications. Compared to alternatives such as Microsoft Word, \LaTeX{} features a whole ecosystem of open-source extensions which are built by an active community.
Many publishers recommend writing scientific papers with \LaTeX{} due to the possibility to supply extensive templates for journals or conference proceedings and get consistent output.

Extension packages for the \LaTeX{} type-setting system are freely distributed over the internet and can be built by anyone with the technical knowledge to do so.
Furthermore, \LaTeX{} as a system relies on text markup to tag the source document with commands which determine the output. This means that it is not necessary to build graphical user interface components to implement new features, such as with WYSIWYG (What You See Is What You Get) word processors.
Implementing the tool as a \LaTeX{} extension is a logical first step, while similar tools for other word processors are also conceivable but require a larger development overhead.

%% Why LuaTeX
Interfering with the standard PDF generation engine (pdfTeX) is not trivial and an extensive task as
it is implemented in the \TeX{} typesetting system, which has many peculiar idiosyncrasies making it very time-consuming to develop new features.
There exists an alternative compiler for LaTeX called \LuaTeX{}\footnote{\url{https://www.luatex.org/}} which features the embedded Lua scripting language and callback hooks to the most important events in the PDF generation process.
Writing the package in \LuaTeX{} allowed to keep the development effort comparatively low and implement the desired functionality quickly.
Implementations for other \TeX{} to PDF compilers are of course still possible in the future if compatibility problems arise.

The developed package SciKGTeX is available on the \TeX{} package archive CTAN\footnote{\url{https://ctan.org/pkg/scikgtex}}
and GitHub\footnote{\url{https://github.com/Christof93/SciKGTeX}}.
Integrating the SciKGTeX functionality into a project can be achieved by downloading the package and putting \verb+\usepackage{scikgtex}+ into the document preamble. To illustrate the configuration of the SciKGTeX package, there is a demo project in the Overleaf service\footnote{\url{https://www.overleaf.com/latex/examples/scikgtex-example/wrhmyrwfgrgw}}. 
For the package to work, it is necessary to compile the \LaTeX{} source with \LuaTeX{}. 
%This can be achieved either by using the \texttt{lualatex} command from the command line or by setting the default compiler of the \LaTeX{} development environment to \LuaLaTeX{}. This option can typically be found in the settings of most modern \LaTeX{} environments.
In the following, we expand on how we implemented the main functionalities depicted in Figure \ref{fig:stories} as a \LaTeX{} package.\\

\textbf{Functionality 1.} Assigning properties to textual entities is implemented by defining new \LaTeX{} commands which can be used to mark expressions in the document.
%% Why these names? Give sentence where they come from.
Five commands were reserved for the most important properties describing a scientific contribution: research problem, objective, method, result, and conclusion.
These command names were chosen from the DEO classes (see \cite{constantin2016}) as suggested by the approach of Oelen et al. \cite{oelen2021}.
The DEO class of \textit{research statement} was adapted to the \textit{research problem} property, which is a central concept of the ORKG vocabulary.
A new property \textit{objective} was introduced to fulfil a requirement identified in the first round of user evaluations.
From this we derive the five predefined commands in SciKGTeX: \verb+\researchproblem{}+, \verb+\objective{}+, \verb+\method{}+, \verb+\result{}+, and \verb+\conclusion{}+. %Additionally, if the paper contains more than one contribution, they can be numbered in the arguments of the marked properties.

A scientific paper typically has a small number of distinct contributions.
In the case that there is more than one contribution in the same document, all the above commands accept an optional argument which allows distinguishing the contributions.
The optional argument can be any identifier, but is most intuitively understood as an enumeration.
For example, annotations \verb+\researchproblem[1]{..}+ and \verb+\researchproblem[2]{..}+ add two separate contributions with respective research problems. If two contributions have a property in common, the property can be assigned to the two contributions using a comma between the arguments, for example, \verb+\method[1,2]{..}+.
This would mean that there are two distinct contributions with the respective research problems which share the same method.

Additional to the 5 mandatory ones, it is also possible to specify other properties with the \texttt{contribution} command.
These other properties can be any arbitrary string in theory but are especially valuable if common interesting properties of scientific subdomains are used. For example, properties of p-value or accuracy are useful for studies that include statistical examinations and can be attached to a contribution with \verb|\contribution{p-value}{0.05}| and \verb|\contribution{accuracy}{0.876}|. In the metadata, these properties will be created as extensions to the ORKG ontology.
If an author wants to reuse a property from another specific ontology, this can be achieved with the commands detailed in functionality 4.\\

\begin{figure}[h]
    \centering
    \includegraphics[width=\linewidth]{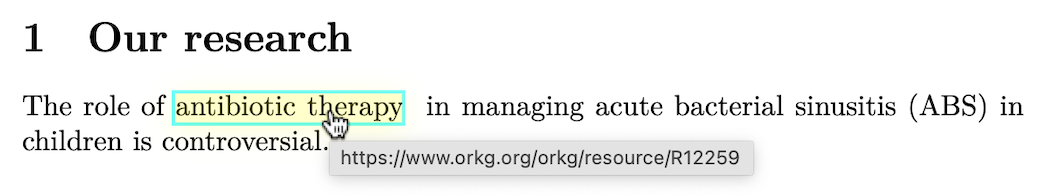}
    \caption{Entity linking rendering}
    \label{fig:hyper}
\end{figure}

\textbf{Functionality 2.} Linking of URIs to textual entities is implemented with the \verb+\uri{}+ command placed inside an annotation. This takes the URI of an entity defined in the web as the first argument and an optional label as the second, see Listing~\ref{lis:1}. If a label is given, it is rendered as a hyperlink to the URI, see Figure~\ref{fig:hyper}.

\begin{lstlisting}[caption=Entity linking, label=lis:1]
The role of \researchproblem{\uri{https://www.orkg.org/orkg/resource/R12259}{antibiotic therapy}} in managing acute bacterial sinusitis (ABS) in children is controversial...
\end{lstlisting}

\textbf{Functionality 3.} Annotation of bibliographic metadata in \LaTeX{} is implemented with the commands \verb+\metatitle+, \verb+\metaauthor+ 
 and \verb+\researchfield+ for the title, authors, and research field respectively, see Listing~\ref{lis:2}. We decided not to use the existing commands in \LaTeX{}, such as \verb+\title+ or \verb+\author+, as they may vary across various templates and require different formatting. The \texttt{meta} prefix is added to the commands to not overwrite the existing ones.

\begin{lstlisting}[caption=Annotated bibliographic metadata, label=lis:2]
\title{\metatitle{Effectiveness of Amoxicillin/Clavulanate Potassium in the Treatment of Acute Bacterial Sinusitis in Children.}}
\author{\metaauthor{Ellen R. Wald} \and \metaauthor{David Nash} \and \metaauthor{Jens Eickhoff}}
\researchfield{pharmacology}
\end{lstlisting}

\textbf{Functionality 4.} The definition of custom properties for annotation is achieved by SciKGTeX with the \verb+\addmetaproperty+ command. This command registers a new namespace for the metadata, which can be given as a first parameter to the command. An abbreviation of this namespace can also be specified and used as a prefix for the annotation. This is useful if ambiguous properties from different ontologies are used. Listing~\ref{lis:3} shows an example.

\begin{lstlisting}[
caption=Custom properties,
label=lis:3]
\addmetaproperty[amo, http://purl.org/spar/amo#]{claim}
\addmetaproperty[patent, https://other.type/of/ontology]{claim}
...\contribution{amo:claim}{The earth is round}.
Our patent has the following claim:
\contribution{patent:claim}{An apparatus to achieve something new.}...
\end{lstlisting}

\textbf{Functionality 5.} The annotation of contributions without rendering them into the document text is implemented with the starred variant of the property commands. For example, having the sentence `the p-value was 0.01\% higher than in the earlier experiment', it may be desirable to report the actual p-value in the metadata instead of the relative change. In such a case, the command can be simply marked with a star (see Listing~\ref{lis:4}). In the rendered sentence, the content of the starred property (0.06) will be invisible.

\begin{lstlisting}[caption=Invisible markup, label=lis:4]
...the p-value was 0.01\% higher \contribution*{p-value}{0.06} than in the earlier experiment...
\end{lstlisting}
%Figure?

\textbf{Functionality 6.} The completeness check of the mandatory metadata properties is implemented through compiler warnings. If any of the five mandatory commands are missing, the user gets a warning in the console. Figure \ref{fig:warn} provides an example of the warning in the Overleaf interface.

\begin{figure}[htb]
    \centering
    \includegraphics[width=\linewidth]{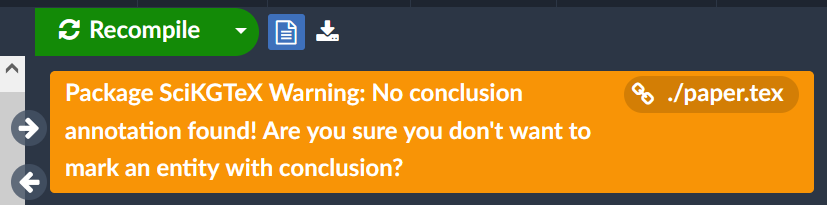}
    \caption{Example of SciKGTeX warning in Overleaf}
    \label{fig:warn}
\end{figure}

\textbf{Functionality 7.} The permanent storage of the annotation is implemented by adding it to the PDF XMP metadata.
The XMP standard is recommended by Adobe as a metadata format for PDF document and is commonly serialized in RDF/XML \cite{xmp-specification16}.
Since RDF is a very fitting format for storage of semantic information, this format can be used to represent the annotations of contributions. In this way, the metadata is merged with the document and can be retrieved by anyone who obtains the PDF.
For the further restricted archival version of the PDF standard called PDF/A the inclusion of XMP metadata is required but also restricted by default to a defined set of properties.
Full PDF/A compatibility can currently only be achieved by using the package's PDF/A compatibility mode, which stores SciKGTeX metadata in a custom catalog entry. This mode also ensures compatibility with other metadata specification package such as hyperxmp or pdfx.

While it is trivial to extract the metadata programmatically from the PDF, manual inspection is not as straight forward, as most PDF viewers are not capable of displaying an arbitrary metadata stream embedded in the document, especially not if it sits in a custom catalog entry.
For this reason, the package produces a corresponding XMP metadata output file for inspection of the created metadata. However, this file is not necessary to distribute the metadata since the whole content is also directly embedded into the produced PDF file in the creation process. As a remark: we also used SciKGTeX to encode the main contribution of this paper.

%Ideally, in the future commonly used PDF viewers will feature possibilities to inspect metadata.

%example of xmp?

\subsection{PDF2ORKG Import Module}

\textbf{Functionality 8.} Annotations from papers annotated with SciKGTeX can be uploaded to the ORKG automatically. This functionality is implemented with the PDF2ORKG import module\footnote{\url{https://github.com/ldrbmrtv/PDF2ORKG}}. The module is written in Python and utilizes the Python ORKG API\footnote{\url{https://orkg.readthedocs.io/en/latest/index.html}}. This module can be integrated into the ORKG interface or into an interface of the paper submission system to automate the ingestion process.

To create a paper instance in the ORKG you need to specify properties such as DOI, title, authors, publication date, publisher, research field, and contributions. At the moment of annotating the paper in \LaTeX{}, DOI, publication date, and venue are unknown, while other properties (title, authors, research field, and contributions) are annotated with SciKGTeX. Table \ref{tab:map} demonstrates the mapping between the ORKG properties and the SciKGTeX annotation.

\begin{table}[htb]
    \centering
    \caption{ORKG properties mapped to SciKGTeX annotations}
    \begin{tabular}{|c|c|}
        \hline
        ORKG property & SciKGTeX command \\
        \hline
        \hline
        DOI & --- \\
        \hline
        Title & \verb+\metatitle+ \\
        \hline
        Authors & \verb+\metaauthor+ \\
        \hline
        Publication date & --- \\
        \hline
        Published in & --- \\
        \hline
        Research field & \verb+\researchfield+ \\
        \hline
        Contributions & \verb+\contribution+ \\
        \hline
    \end{tabular}
    \label{tab:map}
\end{table}

\begin{figure}[h]
    \centering
    \includegraphics[width=\linewidth]{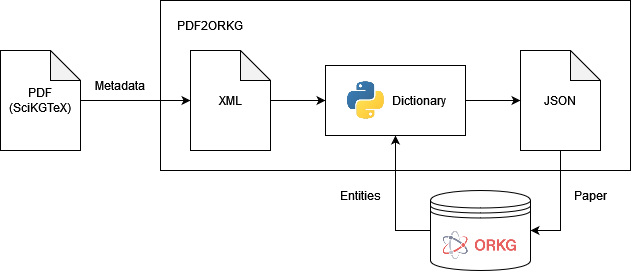}
    \caption{PDF2ORKG dataflow}
    \label{fig:pdf2orkg}
\end{figure}

The data flow of the module for importing the SciKGTeX annotations from PDF files to ORKG is represented in Figure \ref{fig:pdf2orkg}. First, the PDF file is read, and its metadata are extracted in XML format. Then, the module performs HTTP requests to the ORKG API to find the appropriate URIs for the annotated entities and properties. After obtaining all relevant data, a JSON string is formed and passed to the ORKG API method to create or update an instance of the paper. An example of the JSON data is available on the GitHub %repository\footnote{\url{https://github.com/ldrbmrtv/PDF2ORKG/blob/main/metadata.json}}.

Currently, PDF2ORKG imports only the core annotation of the SciKGTeX package, including predefined properties (Functionality 1), and bibliographic metadata (Functionality 3). As ``invisible'' annotation (Functionality 5) results in the same embedded metadata as in Functionality 1, it is also implicitly implemented.
	\section{Evaluation}
\label{sec:evaluation}
The evaluation consists of two parts. In section~\ref{sec:user_evaluation}, we present an evaluation of SciKGTeX with users to investigate the usability of our approach. In section~\ref{sec:proof-of-concept}, we report on a test of the PDF2ORKG import module as a proof-of-concept.

\subsection{SciKGTeX: User Evaluation}
\label{sec:user_evaluation}
The goal of the user evaluation for SciKGTeX was to test the usability of the approach with potential users who are given a small series of tasks to complete using the package.
The evaluation serves to collect a number of metrics which indicate the current usability, and convenience of the developed solution and provide a baseline for further development of the package in the future.
The metrics are designed to reveal more about \hyperlink{RQ1}{research question 1}.
We transformed the research question into three verifiable hypotheses (H1 - H3) which we planned to test in the evaluation.
%H1 and 2 are connected to \hyperlink{RQ1}{RQ 1} and H2 and 3 give insights into \hyperlink{RQ2}{RQ 2}.
%The hypotheses are formulated as alternative hypotheses which we can verify by refuting their negated counterparts, the null hypotheses.
\begin{description}
    \item[H1:] \hypertarget{EH1}{The system is easy and convenient to use.}

    \item[H2:] \hypertarget{EH2}{Annotation of main contributions in a short text summary can be performed in less than 10 minutes.}
    
    \item[H3:] \hypertarget{EH3}{Different annotators produce similar annotations.}
\end{description}

We explored \hyperlink{EH1}{H1} by assessing the perceived usability by the participants of the evaluation. We specifically took into consideration the System Usability Scale (see \cref{sec:metrics}) and feedback from the participants after the evaluation.
\hyperlink{EH2}{H2} sheds light on the simplicity of the annotation tool.
If it holds true, it proves that with little prior training, typical users can learn to achieve the most important objectives of the \LaTeX{} package in little time.
If \hyperlink{EH3}{H3} can be verified, it indicates that the tool is able to produce consistent metadata from different users on different documents. This consistency is a desirable outcome of the tool, since consistent metadata brings many advantages for downstream applications of the data.

\subsubsection{Evaluation Setup}
% candidate selection
For the evaluation, 26 volunteers were recruited from a range of different universities and institutes, mainly in the broader domain of computer science.  %the University of Innsbruck (7), the Leibniz University Hannover (6), the Leibniz Information Centre for Science and Technology University Library\footnote{The Leibniz Information Centre for Science and Technology University Library which is also known as the Technische Informationsbibliothek (TIB) is the German National Library of Science and Technology as well as Architecture, Chemistry, Computer Science, Mathematics and Physics} (5) and several other institutions (8).
%academic backgrounds
A little more than 50\% of the participants (14) were currently pursuing a PhD degree at the time of the test, while another 19\% were master students and 11\% worked as post-doctoral researchers. The remaining 8 participants were working in different research-related positions.

% a priori knowledge
We determined the participants' prior knowledge in \LaTeX{} and semantic web technologies before running the evaluation with a specific set of questions. The exact questionnaire can be found in the experiment artifacts~\cite{Bless.2022}. The familiarity with \LaTeX{} was high among the participants, which adds to the claim that \LaTeX{} is widely used in the academic sector.
The knowledge of Semantic Web concepts was far more varied at a standard deviation of 31 points around the mean on a scale from 0 to 100.
To test the hypotheses, we designed an approximately 30 minutes long evaluation procedure which was executed in a live online meeting with individual participants.
The procedure of the test consisted of the following consecutive steps which were walked through with every participant.\\

\textbf{Evaluation Procedure}
\begin{enumerate}
    \item Give the participant approximately 5 min to read the SciKGTeX documentation\footnote{\url{https://github.com/Christof93/SciKGTeX/blob/main/README.md}
    } and make sure that they understood the idea behind it.
    \item Introduce the participant to the testing environment and give the first task.
    \item Measure time until completion of \hyperlink{task1}{task 1} by the participant.
    \item Introduce tasks \hyperlink{task2}{2} and \hyperlink{task3}{3} and the participant complete them.
    \item Let the participant fill in the survey questionnaire. 
    \item Let the participant give additional oral feedback.
\end{enumerate}

The three tasks we gave to the participants were the following:\newline

\textbf{Evaluation Tasks}
\begin{enumerate}
    \item \hypertarget{task1}{Annotate the 5 properties (background\footnote{\label{fn:background}Background is an old property from the development of the first version of the \LaTeX{} package used for the evaluation. Based on the results of the evaluation, we have revised the 5 predefined properties by removing the background property and adding the new property \textit{objective} (see Section~\ref{sec:package}).}, research problem, method, result, and conclusion) of the main contribution for the given paper~\cite{casillo2022}.}
    
    \item \hypertarget{task2}{Find a unique resource identifier for the term `Natural Language Processing' and link it to the expression in the text. Annotate the resource as a method.}

    \item \hypertarget{task3}{Find a new optional property which you want to annotate in this text. Check if it exists on the ORKG website.}
\end{enumerate}

\subsubsection{Evaluation Metrics}
\label{sec:metrics}
To investigate the hypotheses, we measured three independent variables for each participant in the evaluation setup:
 (i) the System Usability Scale (SUS) score for \hyperlink{EH1}{H1}, (ii) the  time to finish the first evaluation task in seconds for \hyperlink{EH2}{H2} and
 (iii) the Fleiss kappa \cite{fleiss1971} as a measure of agreement between annotators.
The SUS is widely used to measure the usability of software systems \cite{brooke1995}. It is calculated from the user rating of a predefined collection of 10 statements about the user experience of the system.
The Fleiss kappa is a well-known measure to determine the agreement between annotators also known as inter-annotator agreement or inter-rater reliability.
All calculations of metrics can be found in the Python notebook in the experiment artifacts~\cite{Bless.2022}.

\subsubsection{Results}
\label{sec:results}
Below, we  present the results of the evaluation. We provide all the resulting data, including the measured values and user annotations, in the experiment artifacts~\cite{Bless.2022}.\vspace{0.2cm}
% sus score result

\begin{figure}[htbp]
    \centering
    \includegraphics[width=0.48\textwidth]{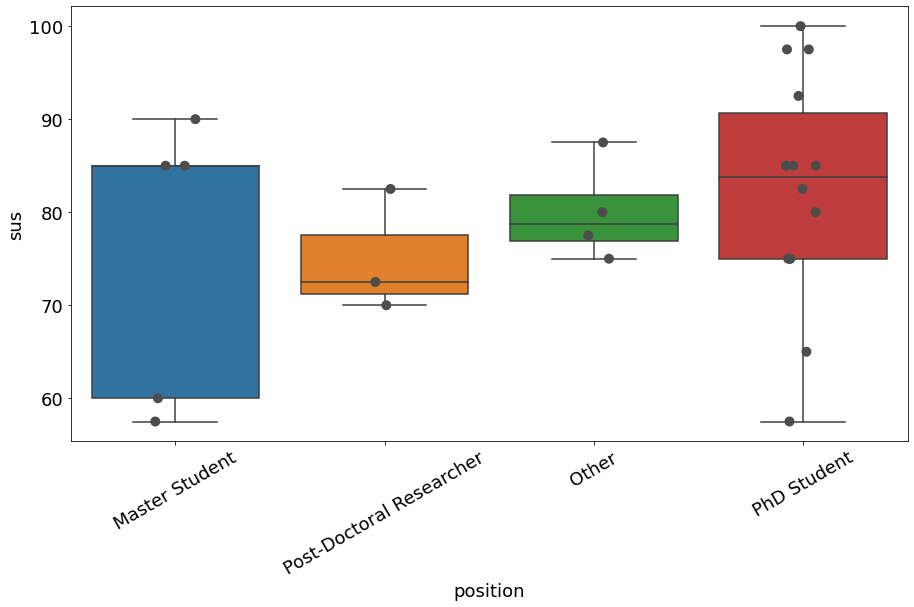}
    \caption{System Usability Scale distributions among the different  groups of participants.}
    \label{fig:sus_score}
\end{figure}

\textbf{System Usability Scale.}
For interpretation of the SUS score, we rely on the work of Bangor et al. \cite{bangor2009} who mapped the percentage-based usability scale to a 7-level adjective scale comprised of `Worst Imaginable', `Bad', `Awful', `Poor', `OK', `Good', `Excellent' and `Best Imaginable'. For the mapping, they ran 212 SUS surveys and asked the adjective ratings alongside the user test. They found that systems rated with `Good' had a mean SUS score of 71.4 ($\sigma = 11.6$ ) while the `Excellent' rating was assigned at a mean score of 85.5 ($\sigma = 10.4$).
The overall mean SUS score of the SciKGTeX package amounts to 79.8 ($\sigma = 11.6$).
The second and third quartiles are situated between 75.0 and 85.0.
This ranks the package clearly closer to `Excellent' than `Good' in terms of matching adjective.
When looking at the different groups of occupations among the participants (see \Cref{fig:sus_score}), it can be observed that PhD students rated a slightly higher mean SUS score than the other groups at 82.3 ($\sigma = 12.34$).
Compared to the PDF annotation tool by Oelen et al.~\cite{oelen2021} (see Section~\ref{sec:related}), SciKGTeX scores \~3 points higher with a similar sample size (23 vs. 26) and standard deviation (11.3 vs. 11.6).

Furthermore, we analyzed the correlation between prior knowledge and the usability score outcome.
%A positive correlation between the \LaTeX{} score and the SUS score for example would suggest that a higher level of prior knowledge of \LaTeX{} leads the user to assign a higher usability score to the package.
To explore the correlation, we calculated Pearson's correlation coefficient between the variables of \LaTeX{} score and SUS score, which amounts to 0.2. This means that there is practically no correlation between these two variables, further implying that prior \LaTeX{} knowledge does not substantially influence the usability of the package.
Also, the Semantic Web knowledge score is not correlated with the SUS score at a Pearson correlation of 0.13.
The combination of near-excellent usability score and independence of prior knowledge makes the package easy and convenient to use, as was hypothesized in \hyperlink{EH1}{hypothesis 1}.\\

\textbf{Annotation Time.}
% time spent result
The variable of time spent on the annotation of the main contribution gives an indication of usability, as it proves or disproves that the idea of the package can be grasped in little time by typical users.
The mean duration of work on task 1 in the evaluation is 7 Minutes 34 seconds at a standard deviation of 3 minutes 21 seconds. This is under our previously defined threshold of 10 minutes and shows that the package can be applied quickly without extensive prior knowledge.\vspace{0.2cm}

\textbf{Inter Annotator Agreement.}
%fleiss kappa result
\Cref{tab:iaa_results} contains the results of the Fleiss kappa inter-annotator agreement values on each of the five property annotations of the main contribution.
It becomes apparent that the first three properties \textit{background}\cref{fn:background}, \textit{research problem} and \textit{method} get far less consistent annotations than the \textit{result} and \textit{conclusion} annotations.
According to Landis et al. \cite{landis1977}, Fleiss kappa values over 0.81 can be considered `almost perfect agreement' whereas 0.61-0.8 is `substantial agreement'.
Applying this to \textit{result} and \textit{conclusion} they can be considered fairly consistent. 
For the other three categories, there are big differences in the text passages which are assigned to them by different annotators.
Manual investigation reveals that there seem to be systematic disagreements on what is considered a \textit{research problem} and a \textit{background}\cref{fn:background} which are often tagged in opposing order. This led us to redefine the \textit{Background} command to \textit{Objective} in the subsequent development cycle of the package as a consequence of the evaluation and user feedback.
The \textit{method} annotation is often split into several annotations of sentence fragments which mention methods, but sometimes the \textit{method} is tagged as a whole block of text.\vspace{1cm}

\begin{table}[htbp]
\caption{Inter-annotator agreement for the different categories of annotations in the \LaTeX{} package.}
\begin{tabular}[c]{|m{0.15\textwidth}|m{0.125\textwidth}|m{0.125\textwidth}|}
\hline
Annotation Type & Fleiss kappa \\ \hhline{|=|=|}

Background & 0.21                                   \\ \hline
Research Problem & 0.44                                   \\ \hline
Method  & 0.24                                   \\ \hline
Result  & 0.74                                   \\ \hline
Conclusion  & 0.81
           \\ \hline

\end{tabular}
\label{tab:iaa_results}
\end{table}

\subsubsection{Threats to Validity}
While the user evaluation is designed to model an actual use case as closely as possible, there are some abstractions which differentiate it from real-world usage. These must be considered threats to the validity of the experiment.

One limitation is that the text which was used as a base for the evaluation tasks~\cite{Bless.2022} was not a document authored by the participants themselves.
Specifically, we chose a randomly selected paper on the topic of Requirements Engineering~\cite{casillo2022}. This topic was selected as most of the candidates which volunteered for the evaluation had at least some background in this research field.
While it was not possible to test each participant on a document that they authored themselves, they should be capable of understanding the text with relative ease to simulate the scenario of self-authorship as closely as possible.
Nonetheless, the fact that the participants are not actually the authors of the text that they annotate in this test compromises the validity of the experiment.
It was not possible to let the participants work on their own texts because the given-above hypotheses could only be evaluated by measuring comparable values in the independent variables, which implied that all participants had to work on the same underlying text.
As a result of this constraint, the group of participants is rather homogeneous in terms of background (mostly computer science) which poses a threat to external validity of the experiment as the results are less generalizable to the whole scientific community. However, a benefit of the more homogeneous group is an increased conclusion validity. Another threat to validity is the relatively small sample size of the experiment.
Subsequent user evaluations should be redesigned to test usability from the author's perspective and comprise a wider base of participants from diverse fields of science, which better represents the targeted user group.
%Additional to that we focused on a homogeneous group of participants (from computer science). in this way, we ensured more consistent results to draw conclusion (improving conclusion validitity). However, as a consequence of increased conclusion validitiy we restricted the generalizablity of our results as a more homogeneous group does not represent the real-world and is therefore a threat to external validitiy.

On the choice of hypotheses, it must be noted that H1 is the only statement which concerns just the technical implementation and functionality of the LaTeX package itself, rather than the broader task of metadata annotation.
Accordingly, hypotheses 2 and 3 are only partly indicators of the usability of the annotation tool.
Testing these hypotheses is also a test of the feasibility of crowd-sourced annotation of contributions since it assesses the difficulty of the tasks which are executed, e.g., finding the main contributions and attributing different properties such as research problem, method, etc. to parts of the text.

\begin{figure*}[htbp]
    \centering
    \includegraphics[width=0.85\linewidth]{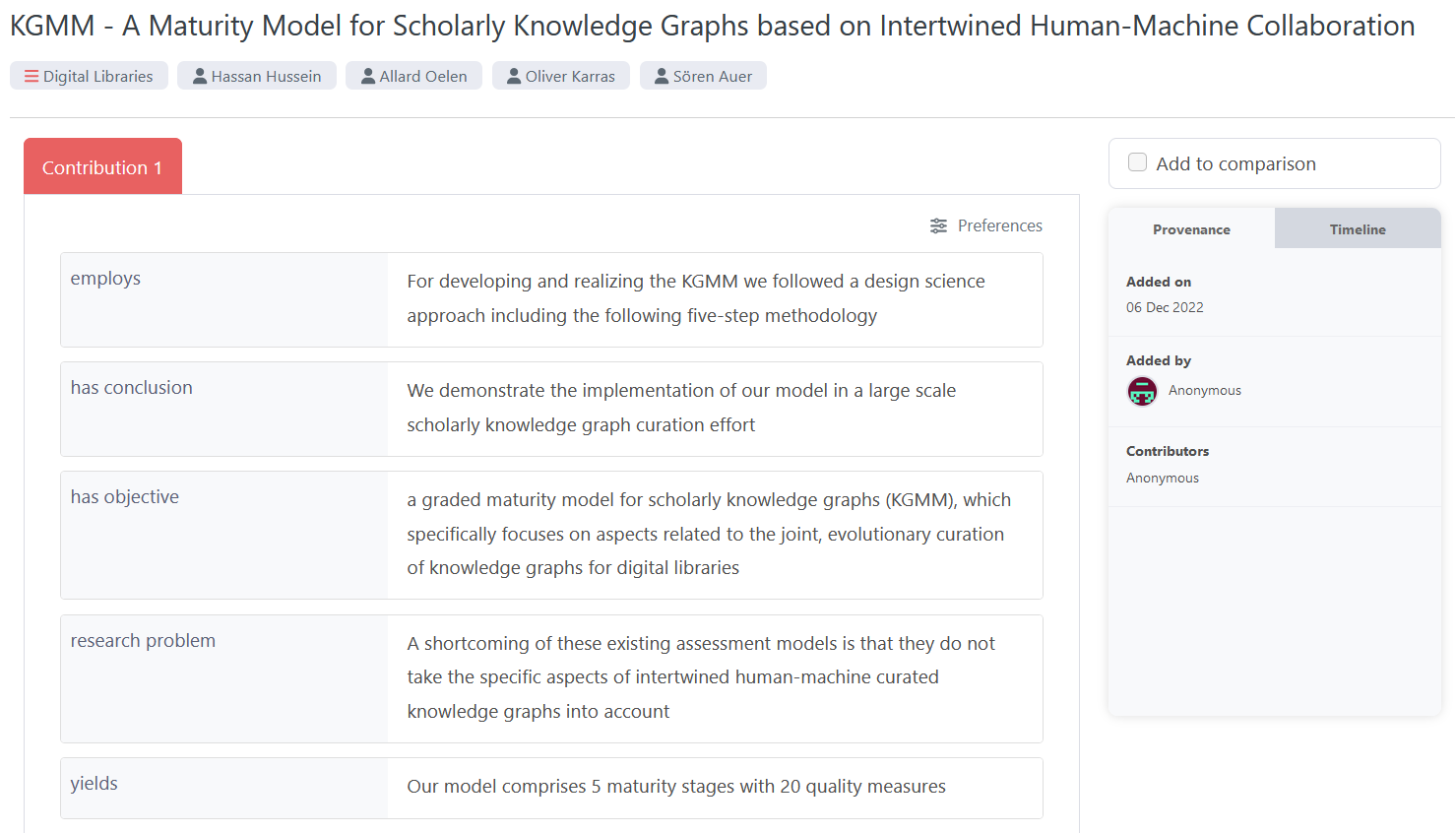}
    \caption{A paper imported to ORKG}
    \label{fig:orkg}
\end{figure*}

\subsection{PDF2ORKG: Proof-of-Concept}
\label{sec:proof-of-concept}

The PDF2ORKG was evaluated by demonstrating the technical feasibility of uploading a paper with its annotations to the ORKG. In particular, we answer \hyperlink{RQ2}{Research question 2} by a technical testing of the PDF2ORKG import module.

%\subsubsection{Testing Setup}
For this test, we asked the first author of a scientific paper~\cite{KGMM} to annotate his paper with SciKGTeX. In this way, we ensured that the process for the proof-of-concept reflects our proposed workflow (see Figure~\ref{fig:idea}). In addition, we tested two uploading modes: 1) Adding a new paper to the ORKG and 2) Updating an existing paper in the ORKG.

%\subsubsection{Results}
When testing the PDF2ORKG module, a paper instance has been successfully uploaded to the ORKG\footnote{\url{https://orkg.org/paper/R282513}} (see Figure~\ref{fig:orkg}). The measurements are provided in Table \ref{tab:time}. It contains the time consumption for the two scenarios, and for the three main steps of the PDF2ORKG workflow. The measurements show that the running time is within tens of seconds, and the main amount of time is spent on uploading data to ORKG. Expectedly, creating a new paper takes longer than updating an existing one.

\begin{table}[htb]
    \centering
    \caption{Time measurements for PDF2ORKG}
    \begin{tabular}{|p{0.4\linewidth}||p{0.15\linewidth}|p{0.2\linewidth}|}
        \hline
        \textbf{Step} & \textbf{Adding paper, s} & \textbf{Updating paper, s} \\
        \hline
        \hline
        Reading and extracting PDF metadata & 0.01 & 0.01 \\
        \hline
        Requesting IDs from ORKG & 0.07 & 0.07 \\
        \hline
        Uploading data to ORKG & 31.3 & 17.07 \\
        \hline
        \textit{\textbf{Total}} & \textit{\textbf{31.38}} & \textit{\textbf{17.15}} \\
        \hline
    \end{tabular}
    \label{tab:time}
\end{table}
	% findings box
	% threads to validity
	
	\section{Discussion}
In this section, we revisit the research questions \hyperlink{RQ1}{RQ 1} and \hyperlink{RQ2}{RQ 2} from \cref{sec:introduction} and discuss the implications of our research.\vspace{0.2cm}
%I provide comprehensive answers to the questions based on the findings presented in this work.

\textit{RQ 1: How can the process of manual semantic annotation of research contributions in scientific articles be simplified?}\vspace{0.2cm}

With the development of the \LaTeX{} package, we have shown that basic semantic information can be directly embedded into the document metadata at the time of document creation, i.e., at the same time as writing the text of the document itself.
Through the usability evaluations, we have shown that the system is understandable, intuitive, and easy to use.
The process of annotation is simple enough for a typical researcher to achieve it in little time.

Different from comparable metadata annotation solutions \cite{oelen2021}, \cite{jaradeh2019}, SciKGTeX is not reliant on any systems other than the \LuaTeX{} document typing system and does not require a connection to the internet to produce the metadata.
This is a simplification compared to the approach where document creation and metadata specification are separated systems. With the novel embedding approach, metadata are directly saved into the PDF files, which saves them from perishing or getting detached from their source material.
Furthermore, the authors themselves dispose of their semantic contribution metadata and do not have to rely on any third-party applications to publish and manage them.

The user evaluation has shown that a representative group of researchers from different universities was able to use the \LaTeX{} package to produce valuable contribution metadata.
Annotating the main contribution in a short text was achieved in well under 10 minutes by the majority of the participants with only little prior exposure to the package documentation.
%The participants had no problem using SciKGTeX with and without extensive prior knowledge of \LaTeX{} or Semantic Web concepts. 
The resulting metadata is machine-actionable and can be used to build large knowledge bases which facilitate various applications from which the researchers can benefit in turn. The low inter-annotator agreement in some property annotations poses a threat to the comparability of the produced metadata, which can be a problem for various applications.
This problem will be addressed in future releases of the package by (i) extending the package documentation with clearer examples and (ii) introducing new properties which leave less space for differing interpretations.\vspace{0.2cm}

\textit{RQ2: How can a scientific knowledge graph automatically
ingest research contributions from document metadata?}\vspace{0.2cm}

With the PDF2ORKG import module, we demonstrate a possible implementation for uploading metadata to a centralized knowledge graph such as the ORKG.

%Ildar+
However, it is only a proof of concept that imports only basic annotation. Currently, the ORKG team is implementing a PDF upload tool that utilizes the SciKGTeX embedded annotation including advanced features such as entity linking, custom properties, and multiple contributions. Similar software could also be integrated into Overleaf, conference submission pages, or paper submission systems such as EasyChair. Another issue with using PDF2ORKG is that at the moment of annotating a paper with SciKGTeX, there is no DOI, publication date, or publisher specified. The most straightforward solution to this is to add the missing data manually after the paper is published.

%Besides implementing the import of the advanced features of the SciKGTeX annotation,
Directions for future work include integrating ORKG API calls directly into SciKGTeX. It would provide the possibility for feedback from the ORKG to an author via the warnings inside the \LaTeX{} editor being used. This feedback can be related to, for example, the ORKG entities, recognized in the annotated text.
	\section{Conclusion}
\label{sec:conclusion}
In this paper, we propose a solution for authors of scientific publications to annotate machine-actionable metadata about the contributions of their publications at the time of writing the manuscript.
The semantic annotations serve to build scientific knowledge graphs, which constitute the future digital record of scholarly publications.
Different from previous solutions, we present a \LaTeX{} package called SciKGTeX which allows directly specifying the metadata at the time of document creation and embedding them into the resulting PDF file.
The metadata can be automatically extracted from the PDF file and uploaded to a scientific knowledge graph, such as the ORKG. %Ildar
This is a simplification compared to an approach where the metadata specification is handled through a separate web interface.

%Further, we conduct a user evaluation with 26 participants to measure the usability, the mean time to complete basic annotations and the inter-annotator agreement of annotations created with the package.
%From the results we conclude that the \LaTeX{} package is convenient and easy-to-use with a mean score of 79.8 on the System Usability Scale and an average time consumption of less than 10 minutes to learn and accomplish the main functionality of the package.
%The inter-annotator agreement of the property annotations is still lacking for the property keywords of \textit{research problem}, \textit{method} and \textit{background} which will be addressed in future work.
In essence, SciKGTeX is a successful implementation of an annotation framework for metadata of scientific contributions.
It is arguably simpler than other approaches with the same objective such as \cite{oelen2021} or \cite{RDFtex} and allows the author to specify metadata directly at the time of document creation.
The user evaluation has confirmed that SciKGTeX can be used by the research community to transform scientific content into machine-actionable metadata.
%Ildar+
Additionally, we have presented an implementation of extracting the metadata from the PDF file and importing it to the ORKG, thereby demonstrating how SciKGTeX complements the existing ecosystem of scientific knowledge graphs.
%Ildar-
Further benefits of the approach are compliance with Semantic Web standards, decentralized information storage and the ability to produce enriched PDF documents.
SciKGTeX has the potential to act as an important building tool for large-scale scientific knowledge graphs and to facilitate the development of supportive applications which elevate the modern research workflow to better standards.

% future work
%- versioning?
%- rework documentation
% Should be the last sentence.
Future plans are to transform the software into a more flexible tool with which the users can define arbitrarily complex facts to add to the document metadata in RDF format while relying on simple building blocks.
Meanwhile, the user experience should stay as simple and elegant as possible.
Ideally, the package can be used as a framework by established publishers to define custom metadata templates which can be used for journals, conference proceedings or other use cases. These templates then allow the aggregation of consistent metadata on papers from the same research area. In this regard, we are pleased to report that the \textit{ing.grid} journal\footnote{\url{https://www.inggrid.org/}} is the first journal that already uses SciKGTeX\footnote{\url{https://www.inggrid.org/news/20/}}. The journal provides an additional branch of its \LaTeX{} template for articles that includes the SciKGTeX package\footnote{\url{https://git.rwth-aachen.de/inggrid/template/-/tree/SciKGTeX}}.

%Furthermore, on online \LaTeX{} writing environments like Overleaf it could be possible to give the users
	
	\section*{Acknowledgement}
	The authors would like to thank the Federal Government and the Heads of Government of the Länder, as well as the Joint Science Conference (GWK), for their funding and support within the framework of the NFDI4Ing consortium. This work was partially funded by the German Research Foundation (DFG) - project number 442146713, by the European Research Council for the project ScienceGRAPH (Grant agreement ID: 819536), and by the TIB – Leibniz Information Centre for Science and Technology.

	\bibliographystyle{ACM-Reference-Format}
	\bibliography{biblio}

\end{document}